\documentclass[12pt]{article}
\usepackage{hyperref}

\usepackage{graphicx}

\begin{document}

\title { Unbound exotic nuclei studied by transfer to the continuum reactions}

\author { G. Blanchon$^{(a)}$, A. Bonaccorso$^{(a)}$ and N. Vinh Mau$^{(b)}$\\
\small $^{(a)}$ Istituto Nazionale di Fisica Nucleare, Sezione di
Pisa, 56100 Pisa, Italy.\\
\small  $^{(b)}$ Institut de
Physique Nucl\'eaire, F-91406 , Orsay Cedex, France }

\maketitle

\begin{abstract}
 In this paper we show that the theory of  transfer reactions from bound  to  continuum states is  well
suited to extract structure information from data obtained by performing   ``spectroscopy in the continuum". 
 The low energy unbound states of nuclei such as 
  $^{10}$Li and $^{5}$He can be analyzed and the neutron-core interaction, necessary to describe
the corresponding borromean nuclei $^{11}$Li and $^{6}$He can be determined in a
semi-phenomenological way. An application to the study of  $^{10}$Li is then discussed and it is shown that
the scattering length for s-states at threshold can be obtained from   the ratio of  experimental
and theoretical cross sections. The scattering single particle states of the
system n+$^{9}$Li are obtained  in a  potential model.  The corresponding 
S-matrix is used to calculate the transfer cross section as a function of the neutron continuum
energy with respect to 
$^{9}$Li. Three different reactions are calculated
$^{9}Li(d,p)^{10}Li$, $^{9}Li(^{9}Be,^{8}Be)^{10}Li$,
$^{9}Li(^{13}C,^{12}C)^{10}Li$, to check the sensitivity of the results to the target
used and in particular to the transfer matching conditions.   Thus  the sensitivity of the structure
information extracted from experimental data on  the reaction
mechanism is assessed.

\end{abstract}
\vspace{1em}

\section{Introduction}
This paper deals with the application of the ``transfer to the continuum method", very well understood for normally
bound nuclei \cite{GSV}, to light unbound nuclei  which recently have attracted much attention \cite{sf}-\cite{bj} in
connection with exotic and halo nuclei. Halo nuclei are very complicated systems to describe. In particular the
accuracy of reaction theories used to extract structure information is a key issue of the field.  From the structure
point of view,  simple semi-phenomenological models have been proposed which exhibit the properties of those nuclei in
terms of one (or two) single nucleon wave functions and which make also easy the calculation of  cross sections for
various reactions initiated by such projectiles
\cite{hs}. One-neutron halo nuclei can be described in a two-body model as a core plus one neutron. All the complexity
of the many-body problem, when two-body correlations are important, can be put in an effective one-body
potential between the extra neutron and the core, that is the Hartree-Fock potential plus the contribution due to the
particle-core vibration couplings. This contribution which is small in normal nuclei is so strong in nuclei such as
$^{11}$Be or $^{10}$Li for example that it might be responsible for an inversion of  1/2$^+$ and 1/2$^-$
states \cite{nic1,gc}. It also induces a strong modification of the wave functions which become mixtures of one-single
nucleon state and  more complicated ones formed of a single nucleon coupled with core vibrations. As a consequence the 
one-nucleon spectroscopic factors are smaller than one. They in turn can be extracted from one-neutron removal cross
sections if one has a good description of the reaction. Then the comparison between theoretical and measured
spectroscopic factors constitutes a strong test of the model.

On the other hand two-neutron halo nuclei such as $^{6}$He, $^{11}$Li  have a two-neutron halo
due to the properties of the single extra neutrons which are unbound in the field of the core, the two neutron pair
being weakly bound due to the neutron-neutron pairing force. In a three-body model  those nuclei are described
as a core plus two neutrons. The properties of core plus one neutron system are essential and the model relies on the
knowledge of angular momentum and parity as well as energies and corresponding neutron-core effective potential,
therefore spectroscopic strength for neutron resonances in the field of the core. Again these information can be
directly obtained from the analysis of one-neutron breakup or transfer cross sections.

The two borromean nuclei that have been studied more extensively so far are $^{6}$He \cite{6he},
and $^{11}$Li \cite{ga}-\cite{bhe}. 
  The two
neutron halo is build on a core which in same cases   is itself a radioactive nucleus (i.e.
$^{9}$Li which is the core of $^{11}$Li). They are  "borromean" since the corresponding (A-1)
nuclei are unbound. 
 Thus $^{5}$He,
$^{10}$Li, as well as $^{13}$Be and $^{16}$B exist only as neutron plus core resonance states and it takes an
extra neutron and its paring energy to finally bind
$^{11}$Li and $^{6}$He. However the two neutron separation energy is
typically  very small (0.3MeV in $^{11}$Li).

The study of unbound systems showing resonances very close to particle threshold is giving rise to the very
interesting field of research that can be defined as  ``spectroscopy in the continuum" \cite{mt} and some of the most
recent applications have been discussed in Refs. \cite{sf}-\cite{bj}. Ideally one would like to study the neutron
elastic scattering at very low energies on the ''core" nuclei. This is however not feasible at the moment as many such
cores, like $^{9}$Li, $^{12}$Be or $^{15}$B are themselves unstable and therefore  they cannot be used as targets.
Other indirect methods instead have been used so far, mainly aiming at the determination of the energy and angular
momentum of the continuum states. This information should help fixing the parameters and form of the neutron-core
interaction.  We remind the reader that the problem of a consistent treatment of the nucleon-nucleus interaction
yielding at the same time bound and unbound states has already been studied for normal nuclei \cite{ms,lip} and it
would be extremely interesting to see how 
 generalizations of such approaches could be obtained from studies of exotic nuclei.

The reactions used so far to study unbound nuclei can be grouped
as: projectile breakup following which the neutron-core coincidences have been recorded and the neutron
energy spectrum relative to the core has been reconstructed \cite{mt},
\cite{zin2}-\cite{cha};  multiparticle transfer reactions \cite{112,1} or just one proton \cite{sf}
transfer. In a few other cases  the neutron transfer from a deuteron \cite{santi} or $^{9}$Be target \cite{santi,bj},
both having low neutron separation energy,   has been  induced   and  the neutron has  undergone a final state
interaction with the projectile of, for example
$^{9}$Li. In this way the
$^{10}$Li resonances have been populated in what can be defined a ``transfer to the continuum reaction"
\cite{bb}-\cite{bb4}. Thus  the neutron-core interaction could be  determined in a way which is
somehow close in spirit to the determination of the optical potential from the elastic scattering of  normal nuclei.

In both the projectile fragmentation or the transfer method the neutron-core interaction that one is trying to
determine appears in the reaction as a "final state" interaction and therefore   reliable
information on its form and on the  values of its parameters can be extracted only if the primary reaction is perfectly
under control from the point of view of the reaction theory. In this paper we argue and show that among the several
methods discussed above to perform spectroscopy in the continuum, the neutron transfer method looks very promising
since the reaction theory exists and has been already tested in many cases \cite{bb}-\cite{mme}. This
has been possible thanks to very accurate and  systematic studies of transfer to the continuum reactions in
normally bound nuclei \cite{GSV},\cite{fg}-\cite{lau}.  We anticipate here that one of the characteristics
of the theory is to allow the consistent use of a different neutron-core interaction at each neutron-core
energy.  This is of basic importance for nuclei such as
$^{10}$Li which have two low lying continuum states with $l$=0 and $l$=1, within an interval of about 0.5 MeV from
threshold, which can be reproduced only by using two very different potentials. The energy and state
dependence of most of the effective nucleon-nucleus interactions is still a challenge in Nuclear Physics
studies. We proceed then to the next session where the basic formalism for the transfer to the continuum
theory is given. Then in Sec. 3 the properties of  $^{10}$Li are resumed.  Sec. 4 contains the discussion of the results
and finally some conclusions and outlook are given in
 Sec. 5.

\section {Transfer to the continuum theory} 

A full description of the treatment of the scattering
equation for a nucleus which decays by single neutron breakup following its
interaction with another nucleus, can be found in Refs. \cite{bb,bb1,jer}. There it was shown that
within the semiclassical approach for the projectile-target relative motion,
 the   cross section differential in $\varepsilon_f$, the final, continuum, neutron energy is 
    \begin{equation}
     {d\sigma\over  {d{{\varepsilon_f}} }}=C^2S
    \int_0^{\infty} d{\bf b_c} {d P_{t}(b_c)\over
d{{\varepsilon_f}}} 
    P_{ct}(b_c), \label{cross} \end{equation}
    (see Eq. (2.3) of \cite{bb4}) and $C^2S$ is the spectroscopic
   factor for the initial single particle orbital.  

The core survival probability $P_{ct}(b_c)=|S_{ct}|^2$ \cite{bb4} in Eq.(\ref{cross}) takes
into account the peripheral nature of the reaction and naturally excludes the possibility of
large overlaps between projectile and target.
    $P_{ct}$
is defined in terms of a  S-matrix function of   the core-target distance of
closest approach $b_c$. A simple parameterization is
$P_{ct}(b_c)=e^{(-\ln 2 exp[(R_s-b_c)/a])}$, where the strong absorption radius $R_s\approx
1.4 
 (A_p^{1/3}+A_t^{1/3}) fm$ is defined as the distance of closest approach for a trajectory that is 50\%
absorbed from the elastic channel and
$a=0.6fm$ is a diffusness parameter.

 Therefore   
according to \cite{bb} the matrix element in the
amplitude for a
   transition from a  nucleon bound state $\psi_i$ in the projectile to a final
continuum state 
$\psi_f$ 
   \begin{equation}A_{fi}={1\over i\hbar}
   \int_{-\infty}^{\infty}dt<\psi_{f} (t)|V({\bf r})|\psi_{i}(t)>,\label{1}\end{equation}
can be reduced to an overlap integral between the
asymptotic parts of the wave functions for the  initial and final  state. Here V is the
    interaction responsible for the neutron transition to the continuum. In the case of a light exotic nucleus
interacting with another light nucleus V(r) is just the neutron-target optical potential  V(r)=U(r)+iW(r),
and the
differential probability with respect to the neutron energy  can be
written as

\begin{eqnarray}{dP_{t}(b_c)\over d{{\varepsilon_f}}}
&=&{1\over 8\pi^3}{m \over \hbar^2 k_f}{1\over
2l_i+1}\Sigma_{m_i}| A_{fi}|^2 \nonumber \\ &\approx& {4\pi\over 
2 k_f^2 }\Sigma_{j_f}(|1- \bar S_{j_f} |^2+1-|\bar S_{j_f} |^2)
(2j_f+1)(1+F_{l_f,l_i,j_f,j_i})B_{l_f,l_i}\nonumber \\&=&
\sigma_{nN} (\varepsilon_f){\cal F},\label{anc}\end{eqnarray}
where $A_{fi}$ is given by Eq.(\ref{1}) and we have averaged over the neutron
initial state.

Equation (\ref{anc}) has a very transparent structure which makes it  suitable to describe the kind of reactions we
are interested in   this paper. In fact the term 
\begin{equation}\sigma_{nN} (\varepsilon_f)={ 4 \pi\over 
2 k_f^2}\Sigma_{j_f}(|1- \bar S_{j_f} |^2+1-|\bar S_{j_f} |^2)
(2j_f+1)\end{equation}
 gives the neutron-nucleus free particle cross section. 
$\bar S_{j_f}$ is the neutron-nucleus optical model S-matrix, which is
calculated for each nucleon final energy in the continuum with an energy
dependent optical model.  The two terms $|1- \bar S_{j_f} |^2$ and $1-|\bar
S_{j_f} |^2$ represent the shape elastic scattering and the absorption
respectively. For the cases described in this paper only the shape elastic term
will contribute, since we will discuss scattering states below the first core
excited state and therefore we will use a real optical potential.

The term
\begin{equation}{\cal F}=(1+F_{l_f,l_i,j_f,j_i})B_{l_f,l_i}\end{equation}
represents what in the theory of final state interactions \cite{joa} has been
called the enhancement factor.
$F_{l_f,l_i,j_f,j_i}$ is an $l$ to $j$ recoupling factor between the angular
momenta of the neutron in the initial and final states. It is also energy
dependent and reflects the spin matching conditions well known for transfer
between bound states \cite{11}-\cite{14}. It is important to notice that our
theory takes properly into account not only the angular momentum dependence of
the final continuum states but also their spin. The term
\begin{equation}B_{l_f,l_i}={1\over 4 \pi}\left [{k_f\over mv^2}\right ]|C_i|^2
{e^{-2\eta b_c}\over 2\eta b_c}M_{l_fl_i}.\label{B}
\end{equation}
 contains the  matching conditions between the
initial and final neutron energies and the relative motion energy per particle   $ mv^2\over 2$  at
the distance of closest approach.                                            
$\eta=\sqrt{k_1^2+\gamma_i^2}$ is the transverse component of the neutron
momentum which is conserved in the neutron transition, $\gamma_i={\sqrt{-2m \varepsilon_i}\over \hbar}$ and
$k_f={\sqrt{2m \varepsilon_f}\over \hbar}$ are the neutron momenta in the
initial and final states and
$k_1={{\varepsilon_f-\varepsilon_i-mv^2/ 2}\over
\hbar v}$ is the parallel component of the neutron
momentum  in the initial state. Also $b_c$ is the core-target
impact parameter,
$C_i$ is the initial state asymptotic normalization constant and $M_{l_fl_i}$
is a factor depending on the angular parts of the initial and final  wave
functions \cite{bb2,bb4}. 

An important characteristic of the present formalism is that the transfer probability
 Eq.(\ref{anc}) contains the factor
$1/k_f^2$ which corresponds to the inverse of the neutron entrance channel flux. It was noticed in Ref.\cite{voy} that
if a  ``transfer to the continuum" formalism does not contain such factor then the model cross sections will always
vanish at zero energy, which is unphysical. Our calculated  cross section instead will have in the case  of a virtual
state of exactly zero energy and  
$l=0$  a divergent-like behavior at zero energy, in accordance to experimental data and to the physical expectations for
a s-state at threshold. It should be also noticed that in the term $B_{l_f,l_i}$  there is a modulating factor ${k_f\over
mv^2}\approx {v_f\over v^2}$ which takes into account the matching between the projectile velocity at the distance of
closest approach $v$ and the neutron final velocity in the continuum  $v_f$.

A particularly interesting case is when the final continuum energy approaches
zero. Then only the $l=0$ partial wave contributes and using
${1\over 4}|1-\bar S_0|^2=\sin^2 \delta_0$ , Eq.(\ref{cross}) becomes very
simple if the write it as differential in the final neutron momentum, in particular if the neutron initial
state is also
$l=0$ and we assume unit spectroscopic factor. In that case the spin coupling factor $(1+F_{l_f,l_i,j_f,j_i})$
and the
$M_{l_fl_i}$ factor are
 independent of energy such that the differential cross section finally reads:

 \begin{equation}
     {d\sigma\over  {dk_f }}=\left (\sin \delta_0\right)^2|C_i|^2\left [{\hbar\over
mv}\right ]^2
    \int_0^{\infty} d{\bf b_c} 
{e^{-2\eta b_c}\over \eta b_c}
   e^{(-\ln 2 exp[(R_s-b_c)/a])}. \label{cross7} \end{equation}
If the LHS of the previous equation is measured experimentally, then
 $\left (\sin \delta_0\right)^2$ can be obtained by doing the ratio between
${d\sigma_{exp}/  {dk_f }}$ and the remaining terms in the RHS of Eq.
(\ref{cross7}),   in the limit of zero energy. 
In fact the above equation is well behaved,  because the only dependence on the neutron energy is
contained in the term ${e^{-2\eta b_c}\over \eta b_c}$,
 where $\eta$ goes to a constant  in the limit of zero
energy. Finally  the scattering length can be obtained from $a_s=- 
\mathrel{\mathop{lim}\limits_{k
\to 0\hspace{.28em}}}{tan\delta_0\over k} $. It is interesting to note the similarity between
Eq.(\ref{cross7}) and the corresponding  formula of  the theory of transfer between bound states 
\begin{equation}
     \sigma( {\varepsilon_f })={\pi \over 2} |C_iC_f|^2\left [{\hbar\over
mv}\right ]^2
    \int_0^{\infty} d{\bf b_c} 
{e^{-2\eta b_c}\over \eta b_c}
   e^{(-\ln 2 exp[(R_s-b_c)/a])} \label{cross3} \end{equation}
as discussed
in
\cite{bb} where it was shown that the term $\left (\sin \delta_0 \right)^2$ after integrating over the final
continuum energy, plays the same role as the asymptotic normalization constant of the final bound state
$C_f^2$.

\section {Application to  $^{10}$Li structure}

 Since the link between reaction theory and structure model is made by the optical potential determining the
S-matrix in Eq.(\ref{anc}), once that the theory has fitted position and shape of the continuum n-nucleus energy 
distribution, what can be deduced are the parameters of a model potential. Therefore we are now going to use such a
model to describe  the properties of $^{10}$Li.  $^{10}$Li is unbound and in its low
energy continuum four  states (two spin doublets) are expected to be present due to the
coupling with the $3/2^-$ p-state of the extra proton in the $^{9}$Li core. The states 
with a total spin of $1^-$ or $2^-$ would be due to the coupling with a neutron in a s-state, while coupling with the
 p-state would give $1^+$ or $2^+$. There is already a rich literature on the subject both from the experimental
\cite{santi} as well as from the theoretical point of view \cite{ga}-\cite{tho}. In particular the best
evidences are in favor of $^{10}Li$ having a $1^-$ ground state due to an s-virtual state
close to the threshold.  Recently a proton pickup experiment $d(^{11}Be,^{3}He)^{10}Li$ 
\cite{sf} has definitely confirmed the earlier hypothesis that the ground state of
$^{10}Li$ is the 2s virtual state and that the $1p_{1/2}$ orbit gives an excited state.

\begin{figure}[ht]
\center
\includegraphics[scale=.45, angle=90]{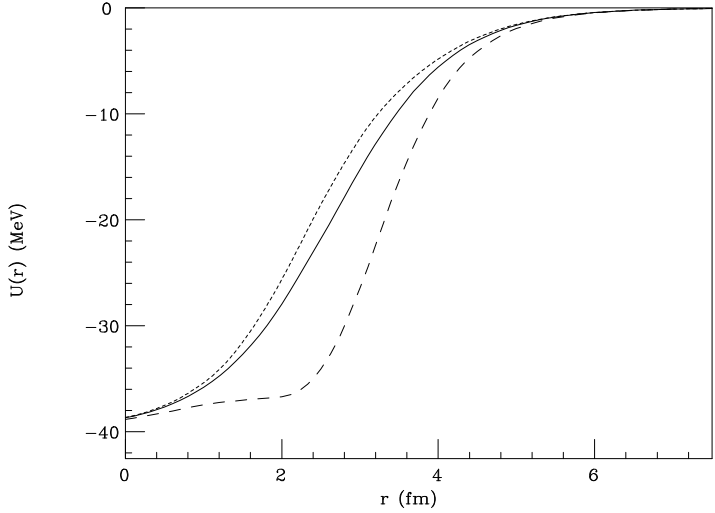}
\caption{\footnotesize Woods-Saxon potential (solid line) and  Woods-Saxon plus surface correction according
 to Eq.(\ref{pot}) 
for the 
$l$=0 state (long dashed line) and the $l$=1 state (short dashed line). }
\label{fig1}
\end{figure}
\begin{table}[h]
\caption{\footnotesize Woods-Saxon and spin-orbit potential parameters.}
\begin{center}
\footnotesize
\begin{tabular}{ccccc}
\hline
\hline
$V_0$      & $r_0$   & $a$    & $V_{so}$ & $a_{so}$ \\
 ($MeV$)   &($fm$) & ($fm$) & ($MeV$)  & ($fm$)   \\
\hline
&&&&\\
-39.83      &1.27   & 0.75   &  7.07    & 0.75     \\
\ \        & \ \   & \ \    & \ \      & \ \      \\
\hline
\hline
\end{tabular}
\end{center}
\end{table}

To describe the valence neutron in $^{10}Li$ we assume that 
the single
neutron hamiltonian with respect to $^{9}$Li has the form 
\begin{equation}h=t+ U\end{equation} where $t$ is the kinetic energy and 
\begin{equation}U (r) =V_{WS}+\delta V   \label{pot} \end{equation}
is the real part of the neutron-core interaction. $V_{WS}$ is  the usual Woods-Saxon potential plus
spin-orbit 
\begin{equation}V_{WS}(r)={V_0\over{1+e^{(r-R)/ a} }}-\left ({\hbar\over m_{\pi}c}\right )^2{V_{so}\over ar}{e^{(r-R)/
a}\over {(1+e^{(r-R)/ a})^2}}{\mathbf  {l  \cdot  \sigma}}
\end{equation}
and $\delta V$ is a correction
 which originates from particle-vibration couplings.  They are important for low
energy states but can be neglected at higher energies. If Bohr and Mottelson collective model is used for the
transition amplitudes between zero and one phonon states, calculation of such couplings suggest the following form
\cite{pvm}:
\begin{equation}
\delta V(r)=16\alpha {e^{{{2(r-R)}/ a}}/({1+e^{{{(r-R)}/ a}}})^4 }\end{equation}
where $R\approx r_0A^{1/3}$. The parameters of $V_{WS}$
for the n-$^{9}$Li interaction used in this paper are those given in Table 1. 
In Table 2 we give the scattering lengths and energy obtained for the 2s and
1p$_{1/2}$ states , with
different values of the strength $\alpha$.

\section {Results and discussion}
 It would be therefore  interesting and important if an experiment could
determine the energies of the  unbound $^{10}$Li states such that the interaction
parameters could be deduced. Two $^{9}Li(d,p)^{10}Li$ experiments have recently been
performed. One at MSU at 20 A.MeV \cite{santi} and the other at the CERN REX-ISOLDE facility
at 2 A.MeV\cite{bj}. For such transfer to the continuum reactions the theory underlined in
Sec. 2 is very accurate. It should be noticed that the theory has usually been applied to projectile breakup
reactions, in order to study single particle excitations in the target. Here it will be applied to single neutron target
breakup leading to excitations of the n-projectile continuum.  
\begin{table}[h]
\begin{center}
\caption{\footnotesize Scattering length
of the s-state, energy and width of unbound p-state and strength parameter for the $\delta V$ potential.}\vskip .1in
\footnotesize
\begin{tabular}{lccccc}
\hline
\hline\
                        &&$\varepsilon_{res}$ &$\Gamma$&$a_s$& $\alpha$\\
 &&(MeV) &(MeV)&(fm)& $(MeV)$\\
\hline
     &${2s_{1/2}}$&  &&323&-12.5\\
&&&&-17.20&-10.0\\
 &${1p_{1/2}}$ & 0.595 &0.48&&3.3\\
\hline
\hline
\end{tabular}
\end{center}
\end{table}

In order to
study the sensitivity of the results on the target, and therefore on the spin selection rules for transfer and on the
energies assumed for the s and p states, we have calculated the reaction
$^{9}Li(X,X-1)^{10}Li$ at 2 A.MeV for three targets d, $^{9}$Be, $^{13}$C. 
 The  $^{13}$C target has been chosen
because in such a case the neutron transfer from the $p_{1/2}$ initial state to the $p_{1/2}$ final state in
$^{10}$Li will be a non spin flip transition
 $j_i=l_i-1/2 \to j_f=l_f-1/2$ while the transfer to the $s_{1/2}$ state
 would be a spin-flip transition which as it is well known  are enhanced at low incident
energy \cite{11}-\cite{14}. $^{14}$N would also be a good target, having a valence
neutron in a   $p_{1/2}$ state, but the absolute cross sections would be smaller as  the
separation energy is larger (10.55 MeV) than in $^{13}$C. It would however provide good matching conditions
at higher beam energies (E$_{inc}\approx 10$ A.MeV). For the other two cases, the initial state is a
$s_{1/2}$ in the deuteron and a $p_{3/2}$ in $^{9}Be$ thus in both cases
$j_i=l_i+1/2$. Then 
 the transfer to the 2s state is a non spin-flip transition which is hindered, while the transfer to the $p_{1/2}$ is
enhanced at low incident energy.  The initial state parameters are given in  Table 3. For each initial state a unit
spectroscopic factor was assumed. 

\begin{table}[h]
 \caption{\footnotesize Targets and  initial state parameters of the bound  neutron.}
\begin{center}\footnotesize 
\begin{tabular}{lccc}
\hline\hline\
   Target& $d$& $^{9}{\rm Be}$ & $^{13}{\rm
C}$\\ \hline
$\varepsilon_i(MeV)$ & -2.22 &-1.66    & -4.95  \\
$l_i$ & 0 & 1 & 1\\
$j_i$ & 1/2 & 3/2 & 1/2\\
$C_i(fm^{-{1\over 2}})$ & 0.95 & 0.68 & 1.88 \\
\hline\hline
\end{tabular}\end{center}
\end{table}
\begin{figure}[ht]
\center
\includegraphics[scale=0.5, angle=90]{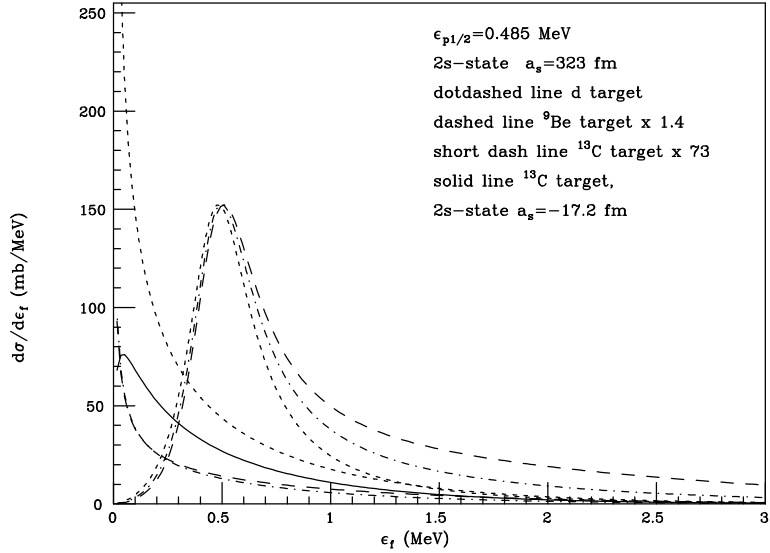}
\caption{\footnotesize Neutron-$^{9}$Li relative energy spectra for transfer to the s and
p continuum states in $^{10}$Li given in Table 2. Dotdashed lines are absolute cross sections for 
transfer from a deuteron target, dashed lines from a $^{9}$Be target, and short dashed line from a $^{13}$C
target. The Be and C cross section have been renormalized to the deuteron cross sections by the factors
indicated on the figure. The solid line is the transfer cross section from the C target to the second s-state
given in Table 2. }
\label{fig2}
\end{figure}

We show in Fig. 2 the neutron energy spectrum relative to 
$^{9}$Li obtained with the interaction and single particle energies of Tables 1 and 2. 
 We define as the resonance energy of the
p-state  the energy at which
$\delta_{j_f}=\pi/2$.  This is also the energy at which
$|1-\bar S_{j_f}|^2$ in Eq.(\ref{anc}) gets its maximum value as it can be seen in Fig. 3.  
 The results of Fig. 2 show that the
peak of the cross section for transfer to the  p-state will determine without ambiguity the position of the
p-state in a target independent way. The measured width instead would depend on the
reaction mechanism, but the ''true" resonance width  can however be obtained from the
phase shift energy variation near  resonance, given by the well known formula 
\cite{joa})
$d\delta_{j_f}/d\varepsilon_f|_{\varepsilon_{res}}=2/\Gamma$, once that the
resonance  energy is fixed.  Using 
this formula we obtained, in the case of the p-state in $^{10}$Li, the value $\Gamma=0.48MeV$ given in Table 2.
From Fig. 2 one can see that the target dependence would influence the extracted width by
about 10-15\%.    It is important to notice that in the approach of this paper the line shape is determined by the 
energy dependence of the phase shift and S-matrix and eventually it could be influenced by an energy dependence of the
potential parameters. Fig. 3 shows indeed the energy dependence of $|1-\bar S_{j_f}|$  for $l$=1. Therefore there
is no need to introduce any a priori form for the resonance shape and width.

 For the s final state
we see that there is a larger probability of population in the spin-flip
reaction initiated by the carbon target. An important question is whether a  measure of the line-shape (or
spectral function) and absolute value of the cross section will determine the
characteristics of the state, and therefore the interaction, also in this case. We have
already shown in Sec.2 that in principle it should be possible. However in order to
elucidate better this difficult question we first remind the reader some of the
peculiarities of the low energy scattering of neutral particles in the
$l=0$ partial wave
\cite{joa,voy,LL}. It is well known that because of the absence of the centrifugal
barrier the energy and width of  an s-state are difficult to define. Therefore we
will in the following study the energy dependence of the phase shift in various
potentials and determine for  each case the values of the scattering length.
 The potential parameters are those of Table 1 and  Table 4. Fig. 4 shows the
behavior, as a function of the neutron momentum, of   $tan\delta_0\over k$ (dotdashed curve) for
the potential (2) of Table 4, of the  cross section (full curve)  calculated in the case of a deuteron target
and of the  factor ${e^{-2\eta b_c}\over \eta b_c}$ (dashed curve) from Eq.(\ref{cross7}). The latter has a
very smooth behavior and therefore it is easy to see that
$(\sin\delta_0)^2$ and then $|a_s|$, could be determined from the ratio between the experimental cross
section and the remaining part of the  RHS of Eq.(\ref{cross7}). 
\begin{table}[h]
\begin{center}
\caption{\footnotesize Strengths of the s-state potential in Eq.(\ref{pot}) and corresponding scattering lengths. Labels
in the first column identify the corresponding curves in Fig. 5.} \vskip .1in
\footnotesize
\begin{tabular}{lcccc}
\hline
\hline\
                      & &$V_0$&$\alpha$&$a_s$ \\ 
& &(MeV)&(MeV)&(fm) \\ \hline
&(1)&-39.83&-4.0&-2.4\\
&(2)&--&-10.0&-17.2\\
&(3)&--&-12.2&-318\\
&(4)&--&-13.5&45.1\\
&(5)&--&-15.0&21.4\\
&(6)&-42.80&-13.3&12.9\\
\hline
\hline
\end{tabular}
\end{center}
\end{table}

\begin{figure}[ht]
\center
\includegraphics[scale=0.5, angle=90]{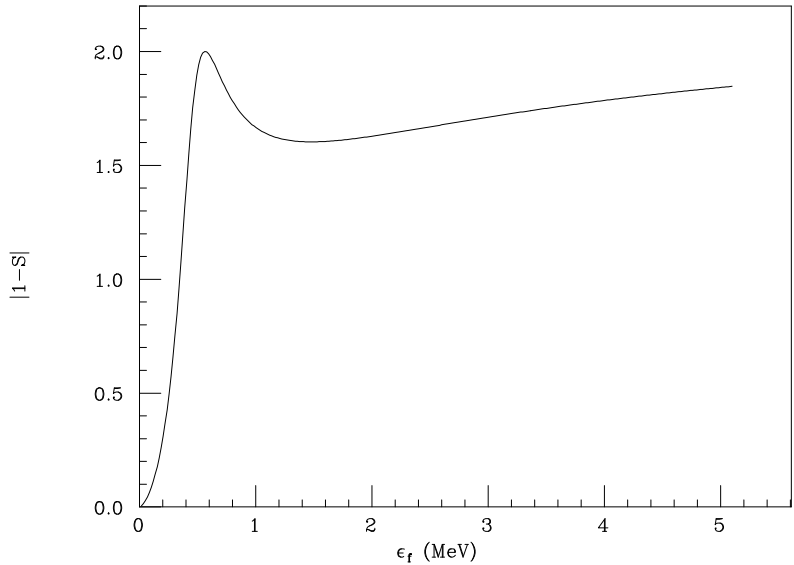}
\caption{ \footnotesize Energy dependence of $|1-\bar S|$  for $l$=1.}
\label{fig3}
\end{figure}

\begin{figure}[ht]
\center
\includegraphics[scale=.5, angle=90]{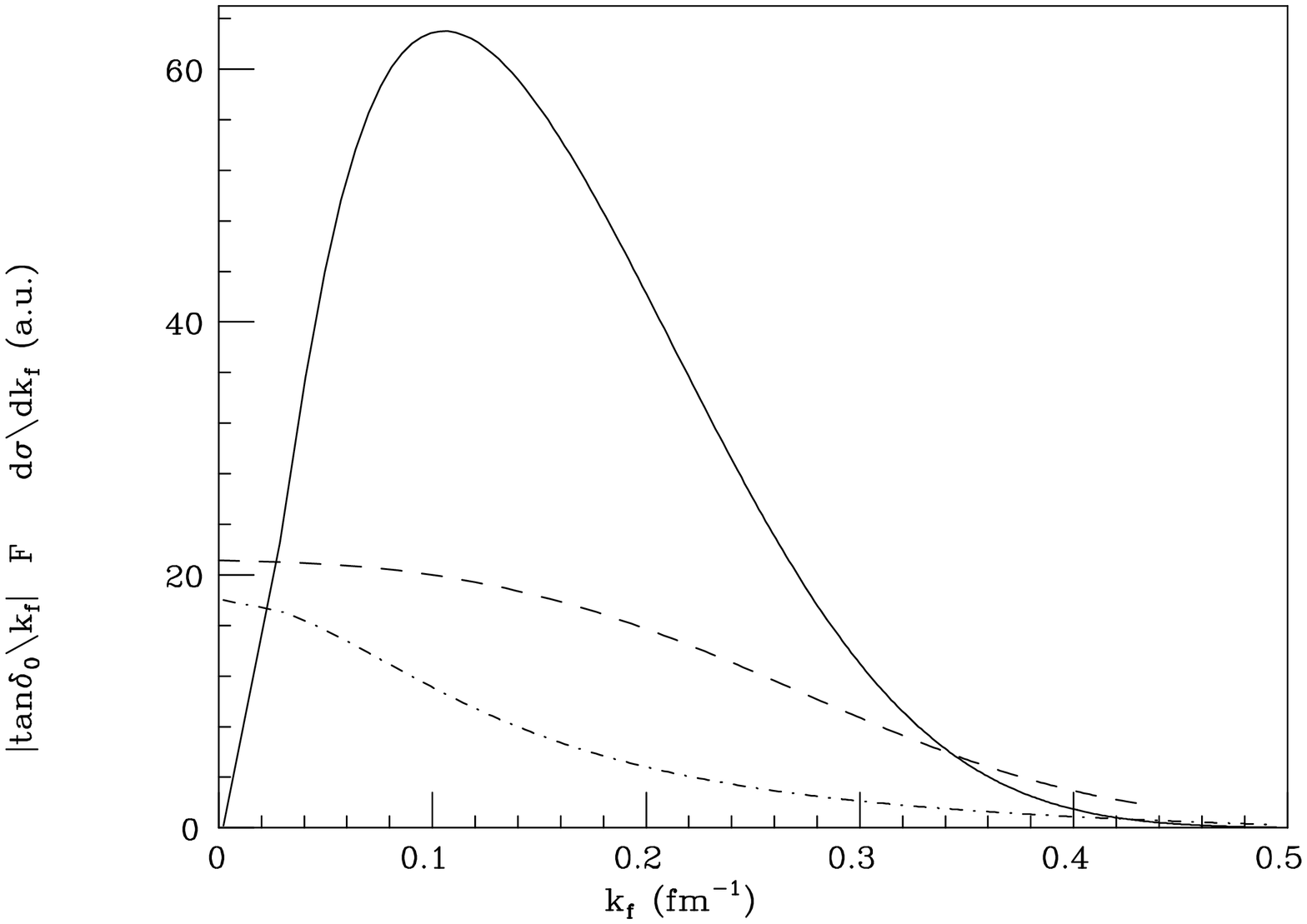}
\caption{\footnotesize $tan\delta_0\over k$ (dotdashed curve) from phase shift for case (2) of
Table 4, factor $F={e^{-2\eta b_c}\over \eta b_c}$ (dashed curve) at $b_c=R_s$ from Eq.(\ref{cross7}) and
cross section (full curve)  for a deuteron target. }
\label{fig5}
\end{figure}
The sensitivity of the  results for transfer to an s-state, on the potential assumed,
is illustrated by Figs. 5a, 5b, 5c, 5d. In Fig. 5a the $l=0$ phase shift is plotted    as a
function of the continuum energy. There are several potentials which give a similar behavior
of the phase shift  but different scattering lengths, (cf. Table 4) and in particular a
different  line shape for the  transfer cross section from a deuteron target  to an s-state, as  shown in
Figs. 5c, 5d. The curves from bottom to top in Fig. 5a, correspond to calculations in the potentials of
Table 4 in increasing order of depth. Therefore the dashed and solid   lines in
Fig. 5c correspond to unbound states with negative scattering lengths, while the long dotdashed 
line  corresponds to a virtual state with a large scattering length consistent with
infinity and therefore of zero energy. Then
 the other three, short dashed, dotted and short dashdotted curves, cases (4), (5) and (6) of Table 4
respectively, correspond to weakly bound states close to threshold.  
  Our results for the phase shifts and scattering length
are consistent with those of the thesis of S. Pita \cite{sf} and with the well known theory of low energy
scattering of neutral particles in s-wave. We have indeed in
Fig. 5a that for unbound states the phase shift is zero at zero energy, then increases up to a maximum value and then
decreases again. Because it never increases going through the value  $\pi/2$
 when the energy increases, as instead it might happen for $l>0$ states, then the states
corresponding to cases (1),(2)  of Table 4 cannot be defined as resonances, even 
though they give rise to an
enhancement of the cross section (see Ref.\cite{joa}, Eq.(4.235) and following discussion). Furthermore
they do not give rise to singularities in the scattering amplitude on the physical sheet of the complex
energy plane. For each of them instead, the scattering amplitude has  a pole at negative energy
$\varepsilon_f=-|\epsilon_{(1,2)}|$ on the un-physical sheet. These poles represent  bound states close to
threshold which give the same free particle scattering cross section as the unbound states, namely
$\sigma=4\pi/(k_f^2+\kappa^2)$ where $\kappa^2=2m|\epsilon|/\hbar^2$ (see Ref.\cite{LL}, Eq. (133.8) and
following discussion). Therefore cases (1),(2)  are broad states with a width of 1-2MeV.  In case (3)
instead the phase shift value is very close to
$\pi/2$ at zero energy corresponding to a virtual state. In fact the S-matrix
gets its maximum value of $|1-\bar S|=2$ in Fig. 5b. Cases (4), (5) and (6) are from potentials which
barely bind  states very close to threshold. The phase shift approaches  the value $\pi$ at zero
energy and the cross sections shown in Fig. 5d are a typical example of how weakly bound states can affect
scattering at low energy. 
\begin{figure}[ht]
\center
\includegraphics[scale=.55, angle=90]{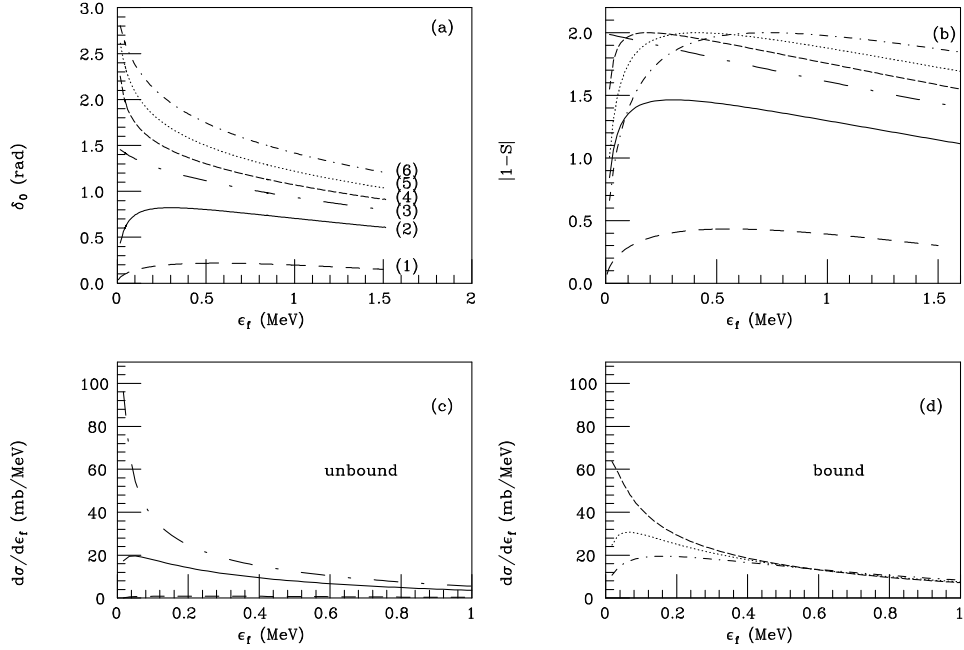}
\caption{\footnotesize Phase shift (a), shape elastic factor $|1-\bar S|$ (b) and cross section (c,d) as a
function of the neutron continuum energy for an s-state and a deuteron target. Figure (c) contains the results
for unbound states with negative scattering length, while  figure (d) for bound states with positive scattering
length.   Labels on the curves identify the corresponding potentials in Table 4. }
\label{fig6}
\end{figure}

On the other hand, what is
clear  is that because of the sharp rise towards zero of the factor
$1/k_f$ and of the less fast decreasing of the
$|1-\bar S|^2$ term of Eq.(\ref{anc}) for $l=0$ , shown in Fig. 5b, the peak of the s-state transfer cross section 
would always be ``downshifted" with respect to the maximum
 of  $|1-\bar
S|$, furthermore a maximum for this term always exists irrespective of the fact that the
s-state at threshold is bound or unbound.   The absolute
value of the corresponding transfer cross sections in  Fig. 5c, 5d increases and has the typical divergent-like
behavior in correspondence to cases (3) and (4) of Table 4.  Then for the more bound states (dotted and
dotdashed line), the transfer cross section decreases again. One such state is  obtained decreasing both the  depth of  the Woods-Saxon and of the surface term and
it corresponds to the smallest positive scattering length in Table 4 ($a_s=12.9fm$).
 The cross section that one would measure in the
continuum, shown in Fig. 5d is just a reflection of the fact that the wave function for a weakly bound s-state has a
long tail and thus some of the transfer strength is in the continuum. In fact, in the region over which the matrix
element in Eq.(\ref{1}) is different from zero, 
the behavior of  bound and unbound state wave functions with energies very close to threshold, is almost the same, due to
the very
large wave length.

Therefore although it would seem quite hard to search experimentally for the energy and ``nature" of weakly bound or
just unbound s-states in exotic nuclei we hope to have shown that  the absolute value of the cross section right at
threshold together with the line shape should determine the scattering length of the state. It appears that
in the case of $^{10}$Li states with  scattering lengths larger than $|a_s|\sim 20fm$ would all lead to a
divergent-like behavior of
$\sigma(\varepsilon_f)$ when $\varepsilon_f \to 0$. The absolute value of $a_s$ can be determined from
the experimental spectrum as discussed in relation to Eq.(\ref{cross7}) and then the parameters of the
n-$^{9}$Li interaction will be fixed as well. Those are the so called virtual states.  One
should also be aware that, as shown in Fig.5d,  resonant-like structures seen in the low energy continuum could be
an indication of weakly bound s-states as well as of unbound s-states. In order to disentangle these two situations
one would obviously need complementary measurements. If the s-state is expected to be the ground state, then the
mass measurement of the nucleus will determine whether it is bound or unbound. In the specific case of 
$^{10}$Li we know indeed that it is unbound. In other cases one could use different targets and/or different
incident energies to study the variation of the maximum of the structures and thus deduce the energy of the
final state from the matching conditions with the initial state.

Finally as the neutron scattering will happen in all partial waves,  if
there is an unbound or virtual s-state the corresponding cross section would seat on top of a background due
to scattering on all partial waves, as one goes away from threshold.  The behavior of such a background would
be different for different potentials and therefore a whole calculation with all relevant partial waves, as
contained in our formula Eq.(\ref{anc}) and a comparison with good resolution data, should help extracting
the correct n-core interaction.  On the other hand it is important to stress that in the case discussed in
this work
 there is no spreading width of the single particle states since the n-$^9$Li interaction is real at such low
energies. In fact the first excited state of $^9$Li is at $E^*=2.7MeV$. For ``normal" nuclei instead the
single particle resonances appear at higher excitation energies (approximately  4-6MeV), and for higher l-values
($l$=6-10). Then it was shown in Ref. \cite{bb6} that the
 spreading width is much larger than the escape width due to the influence of the imaginary part of the potential.

Finally we  conclude that if a transfer to the continuum  experiment could measure with sufficient accuracy (energy
resolution) the line-shapes or energy distribution functions for the s and p-states in $^{10}$Li our
theory would be able to fix accurately the energy of the p-state and the scattering length of the s-state. 
Those in turn could be used to test microscopic models of the n-$^{9}$Li interaction.
The integral  of the energy distribution will
determine the total spectroscopic strength of the state. From our results it appears that such an integral would depend
on the neutron initial state in the target in a way which is however perfectly under control in the theory, since it
is all contained in the B-term given by Eq.(\ref{B}).  
Thus the
spectroscopic strength of the state would be determined by the comparison between measured and calculated values of the
whole energy distribution.

\section{Conclusions and future challenges}

In this paper we have argued that, apart from the experimental difficulties, the transfer to the continuum method is
well suited to study unbound systems such as $^{10}$Li which are the building blocks of borromean nuclei.

There is a very well tested theory to study such reactions, which allows to determine energy distributions for
population of unbound states in absolute value. Provided the same information is available from the experimental
point of view, the theory would allow the determination of the scattering length of s-states and the ''resonance" energy
of unbound single particle states, the associated $l$ and $j$ and the total strength. 
Those studies would eventually be used to determine the neutron-core interaction. 

The great advantage of our method is
that the basic ingredient of the theory is the S-matrix describing the neutron-nucleus scattering. It can be calculated
with an energy dependent potential which can incorporate consistently certain  peculiarities of unbound nuclei  such
as  $^{10}$Li, whose  continuum energy  0-0.5MeV range, for example, contains at least two states with $l$=0,1
obtainable only with two very different potential wells. 

Furthermore the spin-orbit interaction can also be included so
that  at any  energy the contribution from all states with given
$l$ and $j$ can be obtained. 
This is very useful because not only the excitation of states of fixed angular momentum
can be studied, but also the background due to the presence of all other possible angular momentum states can be
calculated and in this way the strength of just one  single particle state can be obtained unambiguously from data
which would contain the contribution from all angular momenta. The theory has the correct behavior when the continuum
energy approaches threshold such that the contribution from virtual states can be distinguished from that from weakly
bound or unbound states.

In this work we have calculated neutron transition probabilities for going from an initial  bound state in a nucleus
to a scattering  state
 including final state interaction with another nucleus. Our  way of describing the final
state interaction in the continuum is through an optical model S-matrix.  A similar approach could be applied to the
treatment of inelastic projectile excitations in which, following its interaction with the target,  a neutron goes
from a bound to  an unbound state with final state interaction in the same nucleus. This is the process which
creates
$^{10}$Li in the final state in the projectile-breakup-type of experiments \cite{mt}. By using such a procedure a very
accurate theory of two neutron breakup could be obtained, incorporating properly the two step mechanism implicit in the
formation of a neutron-core resonance state in reactions like $^{11}{\rm Li+X} \to ^{10}{\rm Li^*+n }\to
^{9}$Li+2n \cite{jl}.

In fact $^{11}$Li breakup and other 2n breakup reactions have often been treated as a process in which the two neutrons
are emitted simultaneously in a single breakup process. This  in principle could be improved by considering  the 
second neutron which decays in flight from a resonant state, as seen for $^{6}$He,  by a breakup form factor
different than that of  the first neutron and by taking into account explicitly the sequential nature of the
process.

{\bf Acknowledgments}

We wish to thank David Brink for several discussions during the preparation of this work.

\end{document}